\documentclass[doublecol,figures]{epl2}

\title{Evolutionary dynamics of cooperation on interdependent networks with Prisoner's Dilemma and Snowdrift Game}
\shorttitle{Evolutionary dynamics on interdependent networks} 

\author{Baokui Wang\inst{1}, Zhenhua Pei\inst{2} \and Long Wang\inst{3}}
\shortauthor{Baokui Wang \etal}

\institute{
  \inst{1} Unit No. 92060 \revision{of PLA}, Dalian, 116041, China\\
  \inst{2} Department of Stomatology, The 306th Hospital of PLA, Beijing, 100101, China\\
  \inst{3} Center for Systems and Control, State Key Laboratory for Turbulence and Complex Systems - Peking University, Beijing, 100871, China
}

\abstract{
The world in which we are living is a huge network of networks and should be described by interdependent networks.
The interdependence between networks significantly affects the evolutionary dynamics of cooperation on them.
Meanwhile, due to the diversity and complexity of social and biological systems, players on different networks may not interact with each other by the same way, which should be described by multiple models in evolutionary game theory, such as the Prisoner's Dilemma and Snowdrift Game.
We therefore study the evolutionary dynamics of cooperation on two interdependent networks playing different games respectively.
We clearly evidence that, with the increment of network interdependence, the evolution of cooperation is dramatically promoted on the network playing Prisoner's Dilemma.
The cooperation level of the network playing Snowdrift Game reduces correspondingly, although it is almost invisible.
In particular, there exists an optimal intermediate region of network interdependence maximizing the growth rate of the evolution of cooperation on the network playing Prisoner's Dilemma.
Remarkably, players contacting with other network have advantage in the evolution of cooperation than the others on the same network.
}

\begin{document}

\maketitle

\section{Introduction}
The problem of cooperation represents a social dilemma characterized by the conflict of interest between group and individuals.
The understanding of emergence and maintenance of cooperation among unrelated individuals in real world is still one of the most challenging problems in social and biological systems.
Evolutionary game theory provides a uniform mathematical framework to deal with this challenge \cite{Alexander1974,Axelrod1981}.
The Prisoner's Dilemma (PD) and Snowdrift Game (SG) are considered as two of the most typical paradigms of investigating cooperation in social dilemmas.
They are both two-person games, where cooperators are prone to exploitation by defectors and the accumulated payoffs of populations are lower than that of pure cooperators\cite{Doebeli2005}.

In conventional forms, both PD and SG are played by two individuals deciding simultaneously whether to cooperate or defect.
Players both receive $R$ by mutual cooperation, whereas mutual defection results in payoff $P$ for both of them.
The highest payoff $T$ is obtained by a player defecting against a cooperator, while the cooperator bearing the cost of $S$.
In the PD, $T>R>P>S$ is achieved.
\revision{Consequently, it is best for individuals to defect regardless of the co-player's decision.
And, defection is the evolutionarily stable strategy (ESS) in well-mixed infinite populations, even though individuals would be better off if they cooperated\cite{Haurt2004}.}
Thus, social dilemma is obvious.
While in the SG, payoffs $P$ and $S$ have a reverse order, $T>R>S>P$, which fundamentally changes the situation and leads to persistence of cooperation.
\revision{As a result,
the replicator dynamics of the SG converges to a mixed stable equilibrium where cooperators and defectors coexist.
In this state, the population payoff is smaller than that of full cooperation, hence the SG still represents a social dilemma\cite{Fu2007PLA}.}

Several mechanisms have been proposed in different contexts to elucidate the ubiquitous cooperative behaviors, such as kin selection, direct reciprocity, indirect reciprocity, network reciprocity and group selection\cite{Nowak2006}.
In addition,
voluntary participation\cite{Szabo2002},
stochastic interactions\cite{Traulsen2006,Chen2008},
social diversity\cite{Santos2008,Perc2008},
\revision{co-evolution\cite{Zimmermann2005,Pacheco2006,Fu2009,Wu2010}},
migration\cite{Cardillo2012},
punishment\cite{Amor2011},
reward\cite{Szolnoki2010},
coordinated investments\cite{Vukov2011},
Matthew effect\cite{Perc2011},
adaptive and bounded investment returns\cite{Chen2012}
and conditional strategies\cite{Szolnoki2012}
have been put under tight study, which provide a good beginning for exploration of the dynamics of cooperation on interdependent networks\cite{Buldyrev2010,Gao2011,Gao2012,Parshani2010,Brummitt2012}.
\revision{Subsequently,
multiplex structure\cite{Gomez-Gardens2012},
biased utility functions\cite{Wang2012EPL},
probabilistic interconnection\cite{Wang2012},
interdependent network reciprocity\cite{Wang2013SR}
and co-evolution\cite{Wang2014NJP,Tang2014PONE}
have been studied intensively to reveal the intrinsic mechanisms of evolutionary dynamics of cooperation on interdependent networks.
}

\revision{
Nevertheless, all the works mentioned above are traditionally modelled as the same games on interdependent networks.
We note that, in social behaviors and biological activities, individuals of distinct regions may engage in different interactions with their local neighbors.}
In particular, strategic information can be shared between regions by information transmission\cite{Szolnoki2013}.
Consequently, this phenomenon should be considered as different game models played respectively on interdependent networks in evolutionary game theory.
Thus, it is interesting to study the evolutionary dynamics of cooperation in a more realistic system constructed by at least two interdependent networks playing different games, such as PD and SG, and far less attention has been paid to this type of system.

\revision{For simplicity and focusing on the interdependence between networks,
we employ two identical spatial structures $A$ and $B$ of the same size, where PD is played on network $A$ and SG no network $B$.}
Individuals on networks $A$ and $B$ can only play with their local nearest neighbors of the same network, where network interdependence cannot influence the payoffs of individuals directly on both networks.
However, in the strategy adoption process, depending on the interdependence between networks with a certain probability $p$, individuals of both networks cannot only learn from their local nearest neighbors, but also from the long-range corresponding one on the other network.
That is, interdependent networks can contact and influence each other by probability $p$, which is described by the network interdependence.
We also assume that, for $p=0$, $A$ has no connection to $B$, where networks $A$ and $B$ are totally separated and cannot influence each other.
In the opposite limit, $p=1$, all the individuals on $A$ and $B$ are completely correlated in order.
For $0<p<1$, the network interdependence between $A$ and $B$ is subject to a binomial distribution.
In this way, individuals on different networks can influence each other and evolve together.
We find that the evolution of cooperation on network $A$ prospers dramatically with increasing $p$.
The impact of network interdependence on $A$ is much stronger than that on $B$.
Surprisingly, there exists an intermediate region of $p$ leading to the maximum growth rate of cooperation level on network $A$.
Moreover, the interdependence between networks fundamentally affects the evolution of cooperation on two networks, where the formation of clusters are changed significantly.
This system may truly reflect the realistic characteristics of multiplicity and diversity in social and biological systems.

\section{Model}
In order to focus explicitly on the impact of interdependence between two interdependent networks and easily compare our results with previous works, networks $A$ and $B$ employed in our work are both two-dimensional $L{\times}L$ square lattices with periodic boundary conditions and von Neumman neighborhoods, which have no empty sites.
In this situation, networks $A$ and $B$ have the same population size, $N_{A} = N_{B}$, and all the players are surrounded by $M=4$ local nearest neighbors.
We assume that PD and SG are played respectively on networks $A$ and $B$.
Initially, players are designated either as a cooperator or a defector with equal probability on interdependent networks.
It is emphasized that, in the game process, players on both networks can only interact with their local nearest neighbors on the same network to obtain their accumulated payoffs.
The payoff matrices for PD and SG can be conveniently rescaled depending on a single parameter $r$ correspondingly.
For the PD, we get

$\left(
  \begin{array}{ccc}
  \label{array1}
  1 & -r \\
  1+r & 0 \
  \end{array}
\right)$.\\
For the SG, we get

$\left(
  \begin{array}{ccc}
  \label{array2}
  1 & 1-r \\
  1+r & 0 \
  \end{array}
\right)$.\\
In this work, $r$ is constrained to the internal [0, 1].
Thus, players on both networks can accumulate their payoffs \revision{$\Pi$} by interacting with their local nearest neighbors on the same network.
While, in the strategy adoption process, the condition is a little more complicated.
Players on both networks cannot only learn from their local nearest neighbors, but also from one corresponding long-range neighbor on the other network with equiprobability, if they are selected with probability $p$ to connect with this corresponding player on the other network.
It is noted that, probability $p$ is used to describe the network interdependence between interdependent networks.
And, each player can only connect with no more than one player of the same position on the other network with probability $p$.
Following the accumulation of payoffs \revision{$\Pi_x$} and \revision{$\Pi_y$}, player $x$, whether from $A$ or $B$, is allowed to learn from a random neighbor $y$.
Note that, as we described above, the random neighbor $y$ maybe a local nearest neighbor, or a corresponding one from the other network if allowed.
Employing the Monte Carlo simulation procedure, player $x$ adopts the strategy of player $y$ with a probability determined by the difference of their payoffs
\revision{\begin{equation}
\label{eq.1}
W_{(x{\leftarrow}y)}=\frac{1}{1+\exp[(\Pi_{x}-\Pi_{y})/{\kappa}]},
\end{equation}}
where characterizes the noise effects in the strategy adoption process.
\revision{$0<{\kappa}\ll1$ implies that the better performing player is readily adopted, whereas large values $\kappa>1$ constitute the weak selection limit.
And, our simulation results are robust for different $\kappa$ when $p>0$.}
Following a previous study, we simply set $\kappa=0.1$ in this work\cite{Szabo1998}, and mainly focus on the impact of interdependence between interdependent networks.

\section{Simulation and analysis}
In the following, we will show the simulation results of the impact of interdependence on the evolution of cooperation on two interdependent networks of size $100\times100$ respectively to avoid finite size effects.
We define that $\rho_{c}$ is the density of cooperators on network.
Thus, $\rho_{ac}$ is the density of cooperators on network $A$, and $\rho_{bc}$ the density of cooperators on network $B$ correspondingly.
In this work, we adopt the synchronous Monte Carlo simulation update manner.
Unless otherwise stated, all the simulation results shown below are required up to $10^{4}$ generations and then sampled by another $10^{3}$ generations.
The results of fractions of cooperators are averaged over $50$ different realizations of initial conditions.


\begin{figure}
\onefigure[scale=0.42]{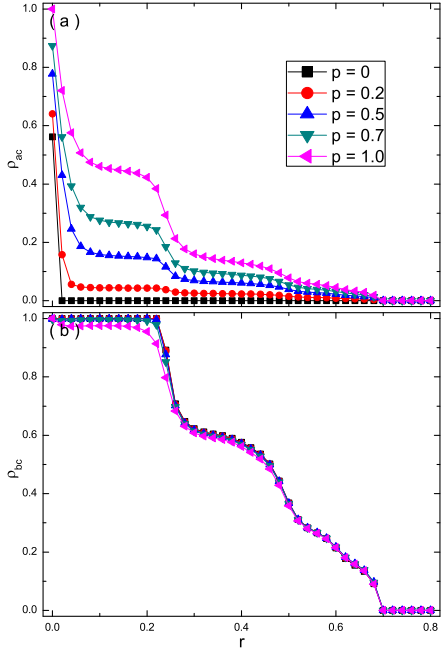}
\caption{
The evolution of cooperation on two interdependent networks as a function of $r$ with different values of $p$.
}
\label{fig1}
\end{figure}

We first study how the network interdependence influences the evolution of cooperation on two networks.
Fig.~\ref{fig1} presents the frequencies of cooperators on interdependent networks as a function of $r$ with different $p$.
Remarkably, as shown in \revision{Fig}.~\ref{fig1}(a), the evolution of cooperation on network $A$ prospers with increasing $p$, which even achieves the full cooperation state when $r=0$ for $p=1.0$.
That is because cooperation is the dominant strategy on network $B$ deducing from our payoff matrices, when $r=0$.
The evolution of cooperation on network $A$ therefore benefits from network $B$ by the network interdependence.
Moreover, comparing with $p=0$, cooperators on network $A$ can remain alive for much bigger $r$ when $p>0$.
As we know, it is best for individuals to defect in the PD regardless of co-player's decision\cite{Haurt2004}.
Thus, the evolution of cooperation can easily vanish in the isolated spatial PD, where $p=0$ in \revision{Fig}.~\ref{fig1}(a).
Nevertheless, with the development of interdependence between networks, the evolution of cooperation on network $A$ is significantly affected by network $B$ which is playing SG.
However, as shown in \revision{Fig}.~\ref{fig1}(b), the cooperation level on network $B$ decreases slightly with increasing $p$.
This phenomenon shows that the evolution of cooperation on network $B$ is also affected by network $A$ through network interdependence between them.
Thus, in this system, the impact of interdependence is mutual on both networks, even though it is much stronger on network $A$ than that on network $B$.

\revision{\begin{figure}
\onefigure[scale=0.45]{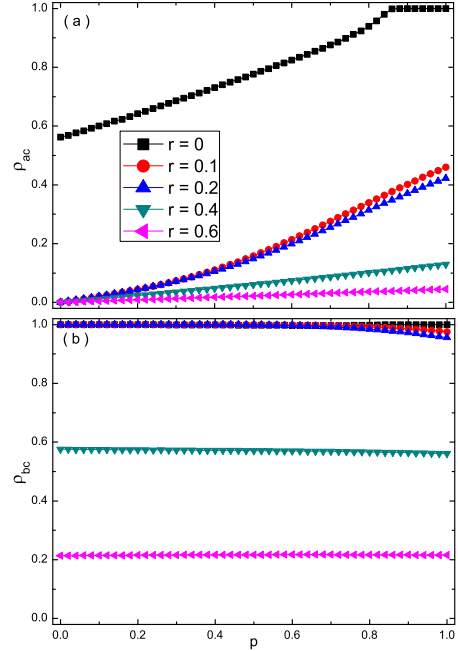}
\caption{
The frequencies of cooperation on two interdependent networks as a function of $p$ with different values of $r$.
}
\label{fig2}
\end{figure}}

To investigate the role of interdependence intuitively, we present the frequencies of cooperation on two networks as a function of $p$ with different $r$ in \revision{Fig}.~\ref{fig2}.
The results shown in \revision{Fig}.~\ref{fig2}(a) clearly evidence that the evolution of cooperation on network $A$ is significantly promoted by the interdependence between networks.
But, the degree of promotion decreases with increasing $r$.
For $r=0$, the cooperation level on network $A$ even achieves the full cooperation state when $p\geq0.88$, which is in accordance with the results shown in \revision{Fig}.~\ref{fig1}(a).
Whereas, if predictable, the evolution of cooperation on network $B$ declines scarcely with different $r$, as shown in \revision{Fig}.~\ref{fig2}(b).
Thus, the impact of network interdependence on the evolution of cooperation on network $B$ is negative and minimal.
Consequently, the impact of network interdependence is inequality on networks $A$ and $B$ due to their different ways of interacting.

\begin{figure}
\onefigure[scale=0.31]{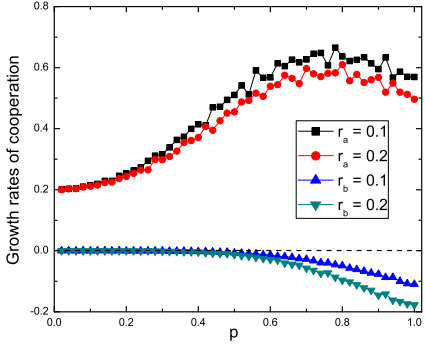}
\caption{
The growth rates of cooperation level on two interdependent networks as a function of $p$ with different $r$.
The black dashed line indicates the base line of coordinates axis.
$r_a$ and $r_b$ indicate parameter $r$ for the evolution of cooperation on networks $A$ and $B$ respectively.
}
\label{fig3}
\end{figure}

In \revision{Fig}.~\ref{fig3}, we define a \revision{quantity} of the growth rate of cooperation $G_c$ to investigate the specific variation of the cooperation level on one individual network with probability $p$.
\begin{equation}
\label{eq.2}
G_c={\mathrm{d}\rho_{c}}/{\mathrm{d}p},
\end{equation}
where $\mathrm{d}\rho_{c}$ is the gradient of $\rho_{c}$, and $\mathrm{d}p$ the gradient of $p$ correspondingly.
Thus, $G_{ac}$ denotes the growth rate of cooperation on network $A$, and $G_{bc}$ the growth rate of cooperation on network $B$ correspondingly.
As shown in \revision{Fig}.~\ref{fig3}, for each fixed $r$, $G_{ac}$ is always beyond the base line with increasing $p$, which is greater than 0.
It means that the existence of network interdependence improves the cooperative behaviors on network $A$, when $p>0$.
However, the results are inverse on network $B$, where $G_{bc}$ is always under the base line for each fixed $r$.
More importantly, there exists an optimal intermediate region of $p$ maximizing the growth rate of the evolution of cooperation on network $A$ evidently.
In other words, there exists an stationary point for the increment of cooperation level on network $A$ when $p>0$, which is around $p=0.8$ for each fixed $r$.
This result effectively proves that although the evolution of cooperation on network $A$ grows with increasing $p$, the growth rate of cooperation level is nonlinear.
And, the promotion of cooperative behaviors is weakened on network $A$ with increasing $r$.
Conversely, the growth rates of the cooperation level on network $B$ are monotonously decreased with different $r$.
And, the decrement of cooperation level on network $B$ is strengthened with increasing $r$.
Thus, the impact of interdependence on the evolution of cooperation is imbalanced on two interdependent networks.

\begin{figure}
\onefigure[scale=0.35]{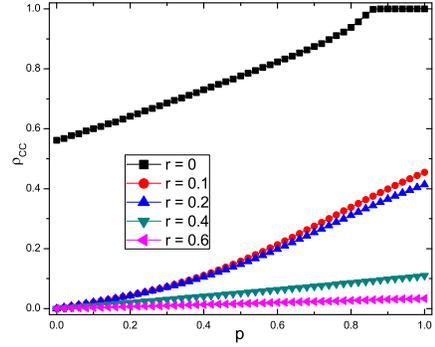}
\caption{
The frequencies of correlated CC strategies between interdependent networks as a function of $p$ with different $r$.
}
\label{fig4}
\end{figure}

In order to investigate the effect of network interdependence intensively, we calculate the frequencies of correlated strategies, such as CC, for corresponding individuals between networks as a function of $p$ with different $r$ in \revision{Fig}.~\ref{fig4}.
Interestingly, as shown in \revision{Fig}.~\ref{fig4}, the frequencies of CC strategies between networks are pretty similar to the evolution of cooperation on network $A$ as shown in \revision{Fig}.~\ref{fig2}(a) for corresponding $r$.
This phenomenon indicates that the interdependence between networks influences the evolution of cooperation on network $A$ directly.
To study the role of correlated CC strategies between networks quantitatively, we employ the simplified correlation coefficient to describe the relationship between CC strategies and network $A$\cite{Wang2012}.
\begin{equation}
\label{eq.3}
r_{ac}=(\rho_{CC}-\rho_{ac}^{2})/(\rho_{ac}-\rho_{ac}^{2}),
\end{equation}
where $\rho_{CC}$ is the fraction of CC strategies between two networks.
We obtain that $r_{ac}$ almost equals to $1$ with different $r$, when $0<\rho_{ac}<1$.
It means that the spreading of cooperative behaviors of individuals on network $A$ is perfectly correlated with the network interdependence between interdependent networks.
In other words, the network interdependence can fundamentally affect the evolutionary dynamics of cooperation on network $A$.
In addition, the results clearly evidence that the interdependence between networks plays similar role on the evolution of cooperation on network $A$ as the probabilistic interconnection between interdependent networks\cite{Wang2012, Wang2013SR}.

\begin{figure}
\onefigure[scale=0.40]{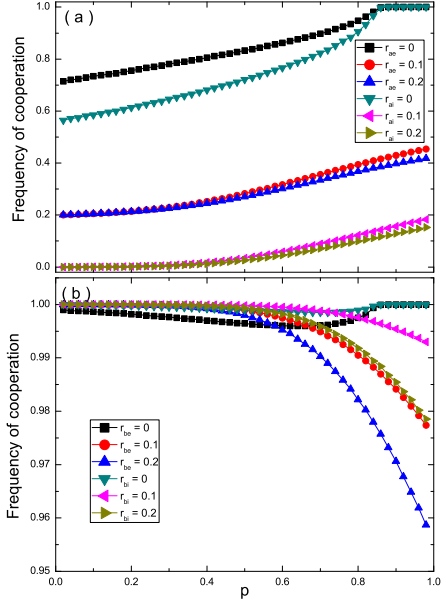}
\caption{
The cooperation level of individuals on interdependent networks have ($r_e$) and do not have ($r_i$) connections with the other network as a function of $p$ with different $r$.
}
\label{fig5}
\end{figure}

We present the frequencies of cooperation of individuals on network $A$ who are selected to build connections with corresponding individuals on network $B$ as a function of $p$ with different $r$ in \revision{Fig}.~\ref{fig5}, to compare with the individuals who can only learn from the local nearest neighbors on the same network.
The results clearly evidence that the cooperation level of individuals contacting with network $B$ is much more prosperous than the other individuals on network $A$, as shown in \revision{Fig}.~\ref{fig5}(a).
This means that these individuals have an advantage in the evolution of cooperation than the other individuals on network $A$ by the interdependence between networks.
However on network $B$, in \revision{Fig}.~\ref{fig5}(b), individuals contacting with network $A$ are on the opposite position, which is at a disadvantage in the evolution of cooperation.
Thus, the impact of network interdependence on the evolution of cooperation on networks $A$ and $B$ is not the win-win situation.
\begin{figure}
\onefigure[scale=0.50]{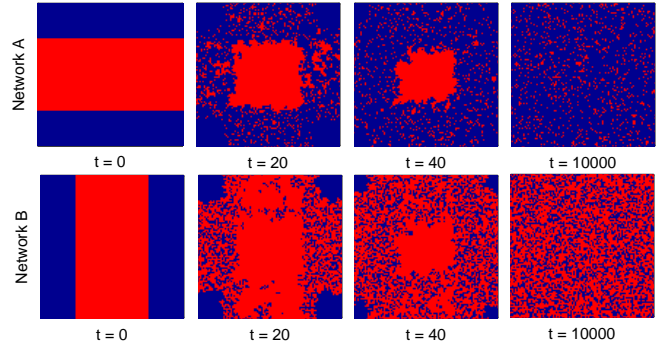}
\caption{
Time evolution of typical distributions of cooperators (red) and defectors (blue) on networks $A$ and $B$ obtained by $r=0.3$ and $p=0.7$ with a prepared initial state.
}
\label{fig6}
\end{figure}

To investigate the evolution of cooperation on interdependent networks intuitively, we present the time evolution of typical distributions of cooperators and defectors on both networks with a prepared initial state in \revision{Fig.~\ref{fig6},
which serves well to highlight two significantly different strategy invasion processes on networks.}
We find that, with the impact of network interdependence, cooperators in the middle domain on network $A$ can survive by forming large and compact clusters initially to reduce the exploitation by defectors.
And, a small number of cooperators appear in the domain of defectors at the beginning phase.
Whereas, the evolution of cooperation on network $B$ prospers by taking advantage of network $A$.
However, with time evolution, cooperators in the middle domain on network $A$ cannot form effective clusters and are gradually eroded by the defectors around, then to reach the steady state.
Ultimately, the evolution of cooperation on network $A$ can only survive by the interdependence between interdependent networks.
On the contrary, although the middle domain of cooperators is gradually eroded by defectors on network $B$, the evolution of cooperation prospers and the domain of cooperators expands to the whole network.
In the steady state, cooperators on network $A$ cannot form large and compact clusters to survive\cite{Nowak1992}.
They have to draw support from network $B$.
The distribution of cooperators on network $A$ is therefore scattered.
Whereas on network $B$, the evolution of cooperation is affected not only by themselves but also by network $A$.
Consequently, the formation of cooperators does not present small filament-like clusters\cite{Haurt2004}, but the intermediate state between the large compact clusters and the small filament-like clusters.
Thus, the interdependence between networks fundamentally influences the evolutionary dynamics of cooperation on both networks simultaneously.

\section{Discussion and conclusion}
We have studied the evolutionary dynamics of cooperation on interdependent networks playing different games, PD and SG.
In this work, two individual networks cannot affect the payoffs of individuals on the other network.
Instead, strategic information is transmitted between networks to affect the evolution of cooperation on both of them.
Coincidentally, Santos\etal explored the evolution of cooperation on interdependent networks with PD and SG\cite{Santos2014}.
\revision{In their work, two networks are modeled as regular random graphs, where individuals on both networks establish intra links with neighbors of the same layer, and inter links with neighbors of the opposite layer.
Biased imitation is introduced to investigate the final level of cooperation reached in each network.}
They showed that as the probability to imitate neighbors from the opposite network increases, the final level of cooperation both networks behaves differently regardless of the population structure.
While in our work, the evolution of cooperation is significantly promoted on the network playing PD with the increment of interdependence between networks.
And, the impact of network interdependence on the network playing SG is negative and slightly.
\revision{It is known that, on an isolated single square lattice, the evolution of cooperation is promoted in PD whereas inhibited in SG, compared with the well-mixed situation\cite{Nowak1992,Haurt2004}.
While in our work, with increasing $p$, the evolution of cooperation is promoted on network $A$ which is playing PD, but inhibited on network $B$ which is playing SG.
This phenomenon clearly evidence that the introduction of interdependence between networks stimulates the promotion of cooperation more conductive in PD, but enhances the inhibition greater in SG.
Moreover, Ohtsuki\etal showed that it is always harder for cooperators to evolve whenever the interaction graph and replacement graph do not coincide\cite{Ohtsuki2007}.
However, the simulation results for PD network in our work are not in line with this, which provides a more general evolutionary dynamics of cooperation on interdependent networks.
}

In addition, there exists an optimal region of intermediate interdependence maximizing the growth rate of the evolution of cooperation on network $A$.
Instead, the growth rate of cooperation level on network $B$ declines monotonously.
Meanwhile, the promotion of the evolution of cooperation on network $A$ is weakened with increasing $r$.
While on network $B$, the evolution of cooperation decreases faster with increasing $r$.
The role of CC strategies of corresponding individuals on both networks have been studied intensively by meas of calculating the frequencies of CC strategies between networks and the correlation coefficient of CC strategies on network $A$.
Remarkably, the frequency of CC strategies is pretty similar to the evolution of cooperation on network $A$ as a function of $p$ with different $r$, where the evolution of cooperation on network $A$ is perfect correlation with the frequency of CC strategies for corresponding $r$.
Then, we have investigated the cooperation level of individuals on both networks contacting with the corresponding individuals on the other network to compare with the individuals playing with local nearest neighbors on the same network.
On network $A$, individuals contacting with the other network have advantages in the evolution of cooperation than other individuals on the same network.
However, it is just the opposite on network $B$.
Thus, the impact of network interdependence on the evolution of cooperation is imbalanced on two networks.
Moreover, the snapshots of evolutionary dynamics of cooperation on two networks have shown that network interdependence fundamentally influences the formation of clusters on both networks.

In summary, the games on interdependent networks studied here are not meant to model a particular real-life situation.
They nevertheless do capture the essence of some situations that are viable in reality.
The evolutionary dynamics of cooperation on interdependent networks with different games may describe the relationship between structured populations with different behaviors.
And, different populations can fundamentally influence each other by sharing information.
By means of this simple model, we would like to reveal the internal mechanisms of how players on interdependent networks affect each other in the evolution of cooperation in real world.
Although this model is simple and cannot include every kind of circumstances existing, we hope this beneficial attempt can highlight the way to explore the evolutionary dynamics of cooperation on interdependent networks.

\acknowledgments
We would like to thank X. J. Chen, Z. H. Yang, J. M. Du and T. Wu for their useful discussion and comments. This work was supported by China Postdoctoral Science Foundation (grant No. 2013M542524).


\begin{thebibliography}{0}

\bibitem{Alexander1974}
  \Name{Alexander R. D.}
  \REVIEW{Annu. Rev. Ecol. Syst}{5}{1974}{325}.

\bibitem{Axelrod1981}
  \Name{Axelrod R. \and Hamilton W. D.}
  \REVIEW{Science}{211}{1981}{1390}.

\bibitem{Doebeli2005}
  \Name{Doebeli M. \and Hauert C.}
  \REVIEW{Ecol. Lett.}{8}{2005}{748}.

\bibitem{Haurt2004}
  \Name{Hauert C. \and Doebeli M.}
  \REVIEW{Nature}{428}{2004}{643}.

\bibitem{Fu2007PLA}
  \Name{Fu F., Chen X.-J., Liu L.-H. \and Wang L.}
  \REVIEW{Phys. Lett. A}{371}{2007}{58}.

\bibitem{Nowak2006}
  \Name{Nowak M. A.}
  \REVIEW{Science}{314}{2006}{1560-1563}.

\bibitem{Szabo2002}
  \Name{Szab\'{o} G. \and Hauert C.}
  \REVIEW{Phys. Rev. E}{66}{2002}{062903}.

\bibitem{Traulsen2006}
  \Name{Traulsen A., Nowak M. A. \and Pacheco J. M.}
  \REVIEW{Phys. Rev. E}{74}{2006}{011909}.

\bibitem{Chen2008}
  \Name{Chen X.-J., Fu F., \and Wang L.}
  \REVIEW{Phys. Rev. E}{78}{2008}{051120}.

\bibitem{Santos2008}
  \Name{Santos F. C., Santos M. D. \and Pacheco J. M.}
  \REVIEW{Nature}{454}{2008}{213}.

\bibitem{Perc2008}
  \Name{Perc M. \and Szolnoki A.}
  \REVIEW{Phys. Rev. E}{77}{2008}{011904}.

\bibitem{Pacheco2006}
  \Name{Pacheco J. M., Traulsen A. \and Nowak M. A.}
  \REVIEW{Phys. Rev. Lett.}{97}{2006}{258103}.

\bibitem{Zimmermann2005}
  \Name{Zimmermann M. G., Egu\'{i}luz V. M.}
  \REVIEW{Phys. Rev. E}{72}{2005}{056118}.

\bibitem{Fu2009}
  \Name{Fu F., Wu T. \and Wang L.}
  \REVIEW{Phys. Rev. E}{79}{2009}{036101}.

\bibitem{Wu2010}
  \Name{Wu B., Zhou D., Fu F., Luo Q.-J , Wang L. \and Traulsen A.}
  \REVIEW{PLoS ONE}{5}{2010}{e11187}.

\bibitem{Cardillo2012}
  \Name{Cardillo A., Meloni S., G\'{o}mez-Garde\~{n}es J. \and Moreno Y.}
  \REVIEW{Phys. Rev. E}{85}{2012}{067101}.

\bibitem{Amor2011}
  \Name{Amor D. R. \and Fort J.}
  \REVIEW{Phys. Rev. E}{84}{2011}{066115}.

\bibitem{Szolnoki2010}
  \Name{Szolnoki A. \and Perc M.}
  \REVIEW{Europhys. Lett.}{92}{2010}{38003}.

\bibitem{Vukov2011}
  \Name{Vukov J., Santos F. C. \and Pacheco J. M.}
  \REVIEW{J. Theor. Biol.}{287}{2011}{37}.

\bibitem{Perc2011}
  \Name{Perc M.}
  \REVIEW{Phys. Rev. E}{84}{2011}{037102}.

\bibitem{Chen2012}
  \Name{Chen X.-J., Liu Y.-K., Wang L. \and Perc M.}
  \REVIEW{PLoS ONE}{7}{2012}{e36895}.

\bibitem{Szolnoki2012}
  \Name{Szolnoki A. \and Perc M.}
  \REVIEW{Phys. Rev. E}{85}{2012}{026104}.

\bibitem{Buldyrev2010}
  \Name{Buldyrev S. V., Parshani R., Paul G., Stanley H. E. \and Havlin S.}
  \REVIEW{Nature}{464}{2010}{1025}.

\bibitem{Gao2011}
  \Name{Gao J.-X., Buldyrev S. V., Havlin S. \and Stanley H. E.}
  \REVIEW{Phys. Rev. Lett.}{107}{2011}{195701}.

\bibitem{Gao2012}
  \Name{Gao J.-X., Buldyrev S. V., Stanley H. E. \and Havlin S.}
  \REVIEW{Nat. Phys.}{8}{2012}{40}.

\bibitem{Parshani2010}
  \Name{Parshani R., Buldyrev S. V. \and Havlin S.}
  \REVIEW{Phys. Rev. Lett.}{105}{2010}{048701}.

\bibitem{Brummitt2012}
  \Name{Brummitt C. D., D'Souza R. M. \and Leicht E. A.}
  \REVIEW{Proc. Nat. Acad. Sci. USA}{111}{2012}{0586109}.

\bibitem{Gomez-Gardens2012}
  \Name{G\'{o}mez-Garde\~{n}es J., Reinares I., Arenas A. \and Flor\'{i}a L. M.}
  \REVIEW{Sci. Rep.}{2}{2012}{620}.

\bibitem{Wang2012EPL}
  \Name{Wang Z., Szolnoki A. \and Perc M.}
  \REVIEW{Europhys. Lett.}{97}{2012}{48001}.

\bibitem{Wang2012}
  \Name{Wang B.-K., Chen X.-J. \and Wang L.}
  \REVIEW{J. Stat. Mech.}{2012}{2012}{P11017}.

\bibitem{Wang2013SR}
  \Name{Wang Z., Szolnoki A. \and Perc M.}
  \REVIEW{Sci. Rep.}{3}{2013}{1183}.

\bibitem{Wang2014NJP}
  \Name{Wang Z., Szolnoki A. \and Perc M.}
  \REVIEW{New J. Phys.}{16}{2014}{033041}.

\bibitem{Tang2014PONE}
  \Name{Tang C.-B., Wang Z. \and Li X.}
  \REVIEW{PLoS ONE}{9}{2014}{e88412}.

\bibitem{Szolnoki2013}
  \Name{Szolnoki A. \and Perc M.}
  \REVIEW{New J. Phys.}{15}{2013}{053010}.

\bibitem{Szabo1998}
 \Name{Szab\'{o} G. \and T\"{o}ke C.}
  \REVIEW{Phys. Rev. E}{58}{1998}{69}.

\bibitem{Nowak1992}
 \Name{Nowak M. A. \and May R. M.}
  \REVIEW{Nature}{359}{1992}{826}.

\bibitem{Vukov2013}
  \Name{Vukov J., Pinheiro F. L., Santos F. C. \and Pacheco J. M.}
  \REVIEW{PLoS Comput. Biol.}{9}{2013}{e1002868}.

\bibitem{Santos2014}
  \Name{Santos M. D., Dorogovtsev S. N. \and Mendes J. F. F.}
  \REVIEW{Sci. Rep.}{4}{2014}{4436}.

\bibitem{Ohtsuki2007}
  \Name{Ohtsuki H., Nowak M. A. \and Pacheco J. M.}
  \REVIEW{Phys. Rev. Lett.}{98}{2007}{108106}.



\end{thebibliography}
\end{document}